\documentclass[sigconf]{acmart}

\AtBeginDocument{%
  \providecommand\BibTeX{{%
    \normalfont B\kern-0.5em{\scshape i\kern-0.25em b}\kern-0.8em\TeX}}}

\copyrightyear{2025} 
\acmYear{2025} 
\setcopyright{rightsretained}
\acmConference[CHI EA '25]{Extended Abstracts of the CHI Conference on Human Factors in Computing Systems}{April 26-May 1, 2025}{Yokohama, Japan}
\acmBooktitle{Extended Abstracts of the CHI Conference on Human Factors in Computing Systems (CHI EA '25), April 26-May 1, 2025, Yokohama, Japan}
\acmDOI{10.1145/3706599.3719931}
\acmISBN{979-8-4007-1395-8/2025/04}

\begin{document}

\title{De-skilling, Cognitive Offloading, and Misplaced Responsibilities: Potential Ironies of AI-Assisted Design}
 
\author{Prakash Shukla}
\affiliation{%
  \institution{Purdue University}
  \city{West Lafayette}
  \state{Indiana}
  \country{USA}
  \postcode{47906}}
\email{shukla37@purdue.edu}
\orcid{0009-0002-7416-1758}

\author{Phuong Ngo Ngoc Bui}
\affiliation{%
  \institution{Purdue University}
  \city{West Lafayette}
  \state{Indiana}
  \country{USA}
  \postcode{47906}}
\email{bui42@purdue.edu}

\author{Sean Levy}
\affiliation{%
  \institution{Purdue University}
  \city{West Lafayette}
  \state{Indiana}
  \country{USA}
  \postcode{47906}}
\email{levy34@purdue.edu}

\author{Maxwell Kowalski}
\affiliation{
  \institution{Purdue University}
  \city{West Lafayette}
  \state{Indiana}
  \country{USA}
  \postcode{47906}}
\email{hkowals@purdue.edu}

\author{Ali Baigelenov}
\affiliation{%
  \institution{Purdue University}
  \city{West Lafayette}
  \state{Indiana}
  \country{USA}
  \postcode{47906}}
\email{abaigele@purdue.edu}

\author{Paul Parsons}
\affiliation{%
  \institution{Purdue University}
  \city{West Lafayette}
  \state{Indiana}
  \country{USA}
  \postcode{47906}}
\email{parsonsp@purdue.edu}

\renewcommand{\shortauthors}{Shukla, et al.}

\begin{abstract}
The rapid adoption of generative AI (GenAI) in design has sparked discussions about its benefits and unintended consequences. While AI is often framed as a tool for enhancing productivity by automating routine tasks, historical research on automation warns of paradoxical effects, such as de-skilling and misplaced responsibilities. To assess UX practitioners' perceptions of AI, we analyzed over 120 articles and discussions from UX-focused subreddits. Our findings indicate that while practitioners express optimism about AI reducing repetitive work and augmenting creativity, they also highlight concerns about over-reliance, cognitive offloading, and the erosion of critical design skills. Drawing from human-automation interaction literature, we discuss how these perspectives align with well-documented automation ironies and function allocation challenges. We argue that UX professionals should critically evaluate AI’s role beyond immediate productivity gains and consider its long-term implications for creative autonomy and expertise. This study contributes empirical insights into practitioners’ perspectives and links them to broader debates on automation in design.

\end{abstract}





\maketitle

\section{Introduction}
The rapid advancement of generative AI (GenAI) has led to significant developments across various creative domains. Large Language Models (LLMs) can now engage in sophisticated conversations, generate text, produce high-quality images and videos, and even create interactive UI mockups for websites and applications. Many widely used UX design tools, such as Figma, Miro, and Framer, are integrating AI into their platforms, further accelerating discussions about the role and impact of AI in design. In the UX design community, practitioners are actively discussing how AI-driven tools will reshape their roles and workflows. Concurrently, researchers are examining the implications of AI for design practices, focusing on both the opportunities and challenges posed by AI integration \cite{stige_artificial_2023, inie_designing_2023, li_user_2024, zimmerman_designing_2009, kalving_where_2024, khan_beyond_2025, olsson_how_2021}. While AI has the potential to automate routine tasks, most creative professionals recognize that human input remains essential to the design process \cite{inie_designing_2023}.

Despite the promise of AI-driven tools in design, their integration into cognitive work introduces complexities that extend beyond simple efficiency gains. Predicting how emerging technologies will reshape professional practice is inherently difficult, as these changes often bring unintended consequences affecting work environments, responsibilities, coordination, and interactions with tools \cite{dekker_envisioned_2002}. History has shown that automation can lead to unexpected challenges, sometimes creating inefficiencies rather than resolving them. Bainbridge \cite{bainbridge_ironies_1983} famously outlined several ironies of automation in process industries, many of which remain relevant to contemporary AI integration \cite{endsley_ironies_2023}. With prior cycles of automation hype and setbacks in mind, we aim to investigate whether UX designers are aware of the potential ironies associated with LLM-driven AI. Recent studies have begun examining the paradoxical effects of GenAI on practitioners, emphasizing how automation may, in some cases, hinder productivity rather than enhance it. For example, Simkute et al. \cite{simkute_ironies_2024} draw on human factors research to analyze the usability challenges of GenAI and propose strategies to mitigate its potential drawbacks. However, there remains an opportunity to build upon decades of research in human-automation interaction to better understand its implications for UX design practitioners.





In this work, we explore the following questions: \textit{What are UX designers' perspectives on GenAI's impact on UX practice? What potential ironies and concerns may arise as AI becomes integrated into the design process?}

To address these questions, we analyze data from two sources: blog posts written by UX practitioners and discussions on UX-related subreddits. Blogs have become a widely used medium for practitioners to share insights and opinions in an accessible and cost-effective manner \cite{herring_weblogs_2005}. Platforms such as Medium, Smashing Magazine, and personal blogs host an expanding community of UX professionals actively contributing to design discourse. Additionally, large corporations engage in these discussions through company web pages, publishing articles authored by design team members and leaders \cite{noauthor_google_2024}. Prior research has demonstrated the value of subreddit analysis in capturing UX practitioners' perspectives \cite{shukla_communication_2024, kou_understanding_2018} . By examining both blogs and subreddit discussions, we gain a broad and diverse view of current attitudes toward AI in UX design. Furthermore, we draw on literature from the history of automation to examine potential ironies and unintended consequences of AI adoption in UX practice. This historical perspective helps contextualize the challenges and complexities of integrating AI into creative and cognitive work.

\section{Background}
\subsection{Integrating GenAI into Design Practices}
The integration of GenAI into design practice offers new opportunities while simultaneously raising important questions about the nature of design work. The possible effects have sparked both enthusiasm and apprehension among practitioners \cite{kalving_where_2024, li_user_2024}. Some discussions have revolved around the potential of GenAI to augment human capabilities and automate mundane tasks \cite{kalving_where_2024, khan_beyond_2025, olsson_how_2021}. However, this integration introduces ethical dilemmas, practical challenges, and questions about the evolving roles of designers \cite{chaudhry_concerns_2024}.

One central discussion revolves around AI as a collaborative partner versus a mere tool. While some designers view AI as a soundboard or co-worker, others insist on maintaining human control over the creative process \cite{kalving_where_2024, khan_beyond_2025}. Concerns exist regarding the potential for AI to diminish the designer's role and erode essential skills, particularly among junior designers \cite{li_user_2024}.

Ethical considerations are significantly important, with discussions focusing on copyright issues, the legitimacy of AI-generated content, and the perpetuation of biases \cite{chaudhry_concerns_2024, kalving_where_2024, knearem_exploring_2023}. The `black-box' nature of AI introduces variability in control, making it challenging for professionals to predict outcomes accurately \cite{law_generative_2025}, raising ethical questions about who maintains responsibility for outputs when processes are not transparent.

A key tension lies in balancing efficiency with creative freedom. While AI can streamline processes such as data collection and prototyping, there is an ongoing debate about whether AI intervention constrains creative exploration \cite{khan_beyond_2025}. For instance, AI-enabled design tools are typically focused on graphical outputs and fail to provide support for other aspects of creativity like those relating to design thinking  \cite{khan_beyond_2025, uusitalo_clay_2024}. Additionally, because AI tools can quickly generate high-fidelity prototypes, processes relating to creativity that occur early in the design process, and typically with more low-fidelity materials, are being cut out \cite{khan_beyond_2025}. Despite noted challenges, several potential roles for AI in supporting design have emerged, including research support, idea generation, and exploration of alternative designs \cite{khan_beyond_2025}. However, designers tend to value AI tools that offer greater control over creative aspects of their work.

Several open challenges and research questions remain: How can designers' sense of control be maintained when using AI tools? How can AI literacy be promoted among designers to ensure responsible and effective AI use? How can the variety of ethical concerns be addressed? How can AI tools be designed to foster reflection and critical thinking during the design process? And How can AI tools better support the iterative and non-linear nature of creative workflows?


\subsection{Parallels with the History of Automation}
The integration of AI into modern design tools and practices is not without precedent. Decades of research on human-automation interaction reveal recurring challenges and unintended consequences that are highly relevant to contemporary AI-driven tools. In this section, we highlight two key areas of prior work that provide critical context for understanding AI’s role in UX practice. First, we discuss the \textit{ironies of automation}, a foundational concept introduced by Bainbridge \cite{bainbridge_ironies_1983}, which describes paradoxical effects of automation that can increase complexity and cognitive demands rather than reducing them. These ironies provide a lens through which to examine how AI might alter UX design work in unexpected ways. Second, we address the \textit{function allocation problem} and the persistent \textit{substitution myth}, which assumes that automation can directly replace human tasks without fundamentally changing system dynamics \cite{dekker_maba-maba_2002}. This perspective is critical for understanding the limitations of AI-driven design tools and the need for a more integrated, cooperative approach to human-AI collaboration.

\subsubsection{Ironies of Automation}
Bainbridge's seminal work on the ironies of automation \cite{bainbridge_ironies_1983} highlighted how automation, intended to simplify tasks and reduce human involvement, can paradoxically make the human operator's role more critical and complex. Automated systems introduce complexities that may necessitate human intervention, particularly during unforeseen scenarios. Here we briefly outline several of the core ironies noted by Bainbridge.

\textit{Deskilling} may occur when reliance on technological systems starts to decrease existing skills or prevents skill development in an individual \cite{sutton_how_2018}. The irony is that operators are expected to monitor and take over control of systems they no longer possess the skills to manage effectively. This has been seen across multiple domains including healthcare, air traffic control, control centers, and more. For instance, a flight crew, after relying on repeated automated flight settings, was unaware that their plane was wandering 70 miles off course, suggesting possible deskilling effects that might arise from technology reliance \cite{mosier_human_1996}. While a core aim of automated industrial processes was to reduce manual workload, it often \textit{increased cognitive workload}. Operators often had to spend more time monitoring automation, interpreting outputs, and diagnosing errors, which ended up being more demanding than the original task. 

Automation can take people `out-of-the-loop,' \textit{reducing situation awareness} and the ability to detect and respond to unexpected events. This can lead to `automation surprises,' where operators are caught off guard by system behavior that deviates from their expectations. The reliability of automation can induce \textit{complacency}, causing operators to become less vigilant and more likely to accept automation outputs without critical evaluation. This over-reliance can lead to errors, especially when the automation fails or provides incorrect information. While automation can make easy tasks easier, it is much more difficult to reduce workload for cognitively demanding tasks, sometimes even making them harder. This kind of `clumsy automation' results in a situation where operators must manage both the automation and the difficult tasks, increasing their overall workload. The \textit{monitoring paradox} refers to the fact that automated control systems are implemented because they can perform tasks better than humans, yet humans are tasked with monitoring their effectiveness. This creates an irony where individuals with potentially diminished skills are expected to oversee complex systems. Finally, automation's efficiency and reliability can disguise operator performance shortcomings. Pre-existing degraded performance can be obscured through automation use, leading to a false sense of security.

The enduring relevance of the ironies of automation is highlighted by its applicability to contemporary challenges presented by AI \cite{endsley_ironies_2023}. While originally focused on traditional automation in industrial and aviation contexts, the core concepts of Bainbridge's work find new expression in the integration of AI systems across various domains. The fundamental premise of Bainbridge's work is that automation, while intended to simplify tasks and reduce human error, often introduces new complexities and critical dependencies on human oversight. This premise holds true for AI systems, which, despite their advanced capabilities, still require human interaction and intervention.

\subsubsection{Function Allocation and the Substitution Myth}
There is a persistent myth in the history of human-technology interaction which posits that novel technologies can seamlessly replace human functions, thereby enhancing system outputs without fundamentally altering the system's structure \cite{lintern_work-focused_2012}. This notion has been referred to as the `substitution myth' \cite{dekker_maba-maba_2002}. This notion is intrinsically linked to the Fitts List \cite{fitts_human_1951}, perhaps the earliest and most well-known attempt to divide responsibilities between humans and machines based on perceived strengths and weaknesses. The Fitts List, also referred to as HABA–MABA ("Humans are better at, Machines are better at"), has been pervasive throughout the history of research on function allocation, although it has faced substantial criticism  \cite{de_winter_why_2014, dekker_maba-maba_2002}.

One key problem with this view lies in the assumption of fixed capabilities for both humans and technology. In reality, automation can engender novel human strengths and limitations in unforeseen ways, thereby undermining the premise of static roles. Capitalizing on automation strengths does not eradicate human limitations; rather, it creates new strengths and limitations, often in ways that are hard to anticipate in advance. Furthermore, the level of granularity for allocating functions is often arbitrary. This can lead to a fragmented understanding of the overall system dynamics and neglect the significance of holistic integration \cite{woods_joint_2006}. The focus on separation and comparability between humans and machines obscures the importance of complementarity and synergistic collaboration.

It is crucial to avoid the pitfalls of this persistent myth by recognizing that automation transforms human practice and compels individuals to adapt their skills and procedures. Designers should prioritize how to foster effective cooperation between humans and AI, rather than attempting to substitute functions based on perceived strengths and limitations. Adopting a systems approach (e.g., see `joint cognitive systems' \cite{woods_joint_2006}) that considers the broader context, including organizational factors, task complexity, and the dynamic interplay between humans and machines, is one important strategy for the future of AI-assisted design.


\section{Method}

Understanding how UX practitioners perceive the role of AI in design requires capturing their discussions in authentic, real-world contexts. Given the evolving nature of AI technologies and the ongoing discourse within the design community, we sought to analyze self-reported insights shared by practitioners in publicly accessible forums. By examining these discussions, we aim to uncover both explicit perspectives and underlying concerns that may not yet be fully articulated in academic literature. To achieve this, we employed a qualitative content analysis approach, drawing from two complementary sources: (1) online articles and blog posts authored by UX practitioners discussing the implications of AI in UX design, and (2) Reddit discussions from UX-focused communities where practitioners openly debate and reflect on AI’s influence on their work. These sources provide a rich dataset that captures both formal reflections and informal, community-driven discourse.

\subsection{Data Collection}
For online articles, we conducted searches using a combination of keywords, including ``UX,'' ``AI,'' ``design,'' and ``LLMs.'' The dataset comprises publicly available articles from platforms such as Medium and community-sourced content websites, specifically focusing on AI technologies and their impact on UX designers. We collected over 120 written works published in the past three years by self-identified UX practitioners and experts in the field. Articles authored by non-UX practitioners and content writers were excluded.

Reddit data was extracted using the PRAW library with keywords such as ``AI,'' ``Generative AI,'' and ``LLMs for AI-related terms, and ``UX'' and ``design'' for UX-related terms. We focused on posts shared in the last three years, resulting in a dataset of 62 posts and 1,575 comments.

\subsection{Data Analysis}
We conducted a hybrid thematic analysis that integrated both inductive and deductive coding methods \cite{fereday_demonstrating_2006}. Our coding process started with a top-down approach that encompassed participants' perceptions on several aspects: the impact of AI on the design process, the future of design practice, limitations of AI, ethical implications, and changing job roles. We followed this with an inductive approach. Four researchers independently engaged in multiple rounds of coding, utilizing both deductive and inductive methods, meeting regularly to discuss. This iterative process led to a consensus on the final codes. Our analysis yielded multiple themes relating to productivity, creativity, human judgment, personalization, and the future of design. However, in this paper, we delved exclusively into themes of productivity, creativity, and human judgment. We then established connections between existing literature on the ironies of AI, the challenges associated with automating manual creative and cognitive work, and our data insights.

\section{Findings}

Our analysis of UX practitioners' discussions on AI in design reveals a nuanced perspective---while there is widespread enthusiasm about AI’s potential to enhance efficiency, concerns persist regarding its impact on creativity, decision-making, and professional expertise. Practitioners see AI as a tool that can streamline repetitive tasks and support ideation, but they also acknowledge risks such as over-reliance, de-skilling, and diminished human judgment in the design process.

Through our thematic analysis, we identified three key themes that characterize how UX practitioners perceive AI’s role in their work: (1) the automation of repetitive processes and its perceived benefits for productivity, (2) AI as a `second brain' that can enhance creativity and ideation, and (3) the continued necessity of human input and judgment in the UX design process. Below, we present these themes in detail, highlighting both the opportunities and tensions that emerge as AI becomes increasingly embedded in UX workflows.

\subsection{Automation of Repetitive Processes}

Our analysis of blog posts reveals that many UX practitioners believe AI will enhance productivity by automating routine design tasks. This belief is based on the idea that AI can handle repetitive work, allowing designers to focus on more strategic and creative aspects of their work. One article states, \textit{``Artificial intelligence (AI) enables UX designers to automate repetitive processes like classifying user activities, forecasting future behaviors, and gleaning pertinent insights from massive amounts of user data. This frees up time for fine-tuning the finished product.''} Similarly, other practitioners express that AI's ability to take over routine tasks enables them to \textit{``get more UX work done''} and \textit{``put greater emphasis on the strategic and creative aspects of their work.''} This perspective aligns with broader industry expectations that AI will optimize workflows by reducing the manual burden of low-level design tasks.



Reddit discussions further support this sentiment, with users highlighting AI’s role in improving daily UX workflows. One user noted, \textit{``Today I asked ChatGPT to create 3 different user flows to brainstorm a customer problem I am working on. Then asked it to create a highly detailed user flow including error states, based on the one I felt was best suited to solve for that problem.''} Another user emphasized the productivity gains AI provides, stating, \textit{``If you're not using AI today on a daily basis in your work, you are missing some significant efficiency and intelligence capabilities that are dramatically reshaping how we do our daily work.''} Additionally, some practitioners view AI as a way to minimize time-consuming, repetitive UX-related tasks, as another user explained: \textit{``I've been looking into ways to use AI tools to be more effective and efficient with my time as a UX professional. I find many UX-related tasks can be time-consuming meta-work or repetitive tasks that just require context-specific adjustments to things like research scripts or UX workflows for usability testing.''}

However, not all practitioners fully endorse the argument that AI saves time. Some express skepticism, questioning whether AI’s productivity benefits outweigh the additional effort required to verify AI-generated outputs. One user remarked, \textit{``I’m starting down this road more and I’m not seeing the value—if it’s AI within Dovetail to help identify themes, that’s one thing, but having to verify feels like the ‘saves time’ argument goes out the window.''} Despite these concerns, the overall sentiment remains largely positive, with many designers embracing AI as a tool for automating repetitive tasks.

\subsection{AI as a ``Second Brain'' supporting Creativity}

Practitioners express varied opinions on AI’s role in creativity. Many believe AI can serve as a \textit{``second brain''}, assisting with ideation and expanding creative possibilities. One participant described this potential, stating, \textit{``Indeed, we should embrace the probabilistic nature of AI, which is one of the main reasons it supports unlimited creativity, leading to the realization that ideation is free with AI.''} Others acknowledge that while AI may aid the creative process, it does not replace human originality, as one participant explained: \textit{``Creativity is based on life experiences [...] AI cannot replace humans because it uses the work that humans create as an input to produce new designs.''} Some also highlight AI’s potential to democratize creativity, making it more accessible to a broader audience. As one participant noted, \textit{``People who may not have had the resources or time to acquire specialized training can now bring their creative visions to life. This democratization means a more diverse range of voices and ideas in every field.''}

Reddit discussions similarly reinforce AI’s role in enhancing creative exploration. One user described AI as a valuable brainstorming tool, stating, \textit{``Embrace the tool and let it help iterate as early drafts. Don’t be narrow-sighted and think this is the only tool you’ll end up using. Take the best ideas it helps you visualize as quicker, more customized sources of inspiration.''} Another user emphasized AI’s ability to support collaborative ideation, saying, \textit{``ChatGPT is good to bounce ideas off, get some info from, and delegate tedious tasks to. Feels like working with a team member who can instantly answer and isn’t busy.''} Additionally, one practitioner described AI as an active collaborator, noting, \textit{``I am treating ChatGPT as a junior UX designer. I have constant conversations on ideas and data that we both collect, and we design together.''}

\subsection{Human Input and Judgment Remains Necessary}

Despite AI’s perceived benefits, UX practitioners overwhelmingly agree that AI cannot fully replace human cognition, but rather serves as a complement to human expertise. One participant emphasized this point: \textit{``While AI can certainly assist with certain aspects of UX design, such as data analysis and pattern recognition, it’s improbable that it’ll completely replace the ability of designers to understand and empathize with users.''} Another participant echoed this sentiment, stating, \textit{``AI won’t completely replace human ingenuity; it will complement human potential.''} Many practitioners highlight the importance of human judgment in the design process, with one noting, \textit{``Design, after all, is not purely a mechanical process. It requires intuition, emotional intelligence, and a deep understanding of human nature, attributes that AI hasn’t yet fully mastered.''} Others warn of AI's limitations, such as its tendency to generate misleading information, as one participant observed: \textit{``Catch `hallucinations,' where the AI makes false assertions with great confidence. As long as the AI’s output is subjected to human review, hallucinations will not damage your results, but you must carefully watch out for them.''}

Reddit users reinforce this view, stressing that foundational UX knowledge is necessary for effectively using AI tools. One user stated, \textit{``If you don't have a base understanding of what it is you’re even asking for or creating, then AI won’t really be all that helpful. You still need someone to troubleshoot and answer questions. And even present the work.''} Another user highlighted AI’s lack of intentionality, explaining, \textit{``If a human pulls together references and creates something new, there’s intention behind it. That intention is the context and relevance that AI has no way to produce algorithmically.''} Another practitioner pointed out the importance of defending design choices, cautioning, \textit{``In order to do that, you need to provide adequate rationale to your decision-making (or lack thereof) to have a healthy debate. If you just say, ‘because AI said so!’ you’ll get smoked!''} Concerns about bias were also raised, with one user warning, \textit{``Without the structured approach of a trained UX researcher, there’s a significant risk of introducing bias at various stages of the process, from question framing to interpretation of results.''} Another user reinforced the necessity of human oversight, stating, \textit{``Eventually, automation driven by systems like this may help make some rudimentary tasks faster, but sound judgment will still be required to evaluate the outputs and adapt them for the context.''}

\section{Discussion}
Our findings highlight both enthusiasm and concern regarding AI in UX design. While AI enhances efficiency and creativity, practitioners also raise issues of de-skilling, cognitive offloading, and shifting responsibilities—echoing historical automation challenges \cite{bainbridge_ironies_1983, woods_joint_2006}. To contextualize these trade-offs, we discuss two key themes: (1) the allocation of responsibilities between designers and AI, and (2) the risks of de-skilling and cognitive offloading. By linking contemporary AI integration to past automation lessons, we emphasize the need to ensure that AI supports---rather than replaces---human creativity and expertise through adequate attention to this history.


\subsection{Designer and AI Function Allocation}

A key finding from our analysis is that many UX practitioners anticipate AI taking over routine tasks such as \textit{``help with user research, prototyping, and usability testing tasks,''} thereby enabling designers to focus on more strategic aspects of their work. This sentiment aligns with findings from Inie et al. \cite{inie_designing_2023}, where creative professionals similarly viewed AI as a means to \textit{``automate the ‘boring tasks'.''} This enthusiasm mirrors historical patterns observed in industrial process control, where automation was initially expected to relieve humans of tedious responsibilities. However, it is important to be cautious of the oversimplification of AI's benefits and the potential for unforeseen consequences. Researchers and developers of UX tools should not fall prey to the substitution myth, assuming that AI can simply replace human labor without altering the nature of the work itself. In reality, new technologies introduce tradeoffs, constraints, and affordances that reshape professional roles, requiring new skill sets and modes of interaction. For instance, considering the ironies discusses above, there is a risk with AI-assisted design that designers will lose visibility into the underlying logic and constraints of AI-generated outputs. For example, platforms like Framer \cite{noauthor_ai_2024} now allow users to generate entire web pages from textual prompts, raising concerns that designers may become disengaged from the rationale behind design decisions. If AI continues to assume greater responsibility for prototyping and ideation, practitioners may struggle to critically evaluate, refine, or justify AI-generated outputs—potentially leading to a diminished role in the creative process.

In the original `ironies of automation' work \cite{bainbridge_ironies_1983}, two important ironies were identified that relate to the intention behind integrating automation into existing practices. Automation is often desired because humans are viewed as unreliable and inefficient, and the more they can be removed from the system the better. This attitude gives rise to two ironies. The first being that errors in designing and developing the automation can become a significant source of operational problems. The second irony is that the humans are often left to do the tasks that the developer could not figure out how to automate, resulting in an arbitrary collection of responsibilities without adequate support. If these ironies hold true for AI-assisted design, practitioners may find themselves tasked with troubleshooting AI-generated outputs rather than engaging in the creative decision-making process itself. Instead of eliminating inefficiencies, AI could inadvertently introduce new forms of cognitive burden, requiring designers to interpret, validate, and correct AI-generated work. As AI continues to integrate into UX practice, it will be crucial to ensure that designers remain actively engaged in shaping design outcomes rather than merely overseeing automated outputs.

\subsection{De-skilling and Cognitive Offloading}

 The potential shift of cognitive tasks like brainstorming and problem framing to AI introduces important questions about the long-term effects on designers' expertise. As AI capabilities advance, it is critical to assess whether increased reliance on AI may lead to unintended skill erosion. Bainbridge \cite{bainbridge_ironies_1983} observed that automation often results in `deskilling,' as workers transition from active participation to passive monitoring roles. While the context of industrial automation differs from that of UX design in terms of process monitoring, similar risks apply. The more that AI assumes responsibility for fundamental design tasks, the less exposure designers may have to the cognitive processes underlying their craft. Our findings suggest that while AI-driven automation is perceived as a means of increasing efficiency, excessive delegation may unintentionally hinder skill development. 

Future work should carefully consider how the ironies and paradoxes identified in the human-automation literature might apply to the integration of GenAI into design. For instance, one potential irony of AI-assisted design is that the reduced time required to generated design artifacts may actually lead to reduced problem solving ability, due to a reduction in incubation periods, which are crucial for creative problem-solving. Researchers have long emphasized the importance of incubation in fostering serendipitous insights, allowing designers to process ideas unconsciously before arriving at innovative solutions \cite{tsenn_effects_2014,ritter_creativityunconscious_2014}. If AI accelerates or bypasses the early stages of ideation, designers may have fewer opportunities for deep reflection and divergent thinking. Additionally, foundational design activities such as sketching and wireframing serve as external cognition aids, facilitating idea development through tangible representation \cite{goldschmidt_dialectics_1991}. If these activities are increasingly offloaded to AI, designers may engage in fewer exploratory iterations, potentially leading to a decline in their ability to conceptualize and refine ideas.


\section{Limitations and Future Research}

While this study provides valuable insights into UX practitioners’ perspectives on AI, several limitations should be acknowledged. First, our analysis is based on publicly available content, meaning our dataset consists of self-reported opinions rather than direct interviews or controlled studies. As a result, we cannot probe participants for clarification or further context, limiting our ability to assess their depth of understanding regarding AI’s impact on UX practice. Second, some of the articles, Reddit posts, and comments included in our dataset may have been AI-generated or influenced by biases inherent in self-reported data. Given the increasing prevalence of AI-generated content online, distinguishing between human and AI-authored discourse remains a challenge and could impact the reliability of our findings. Third, our data reflects perceptions of AI at a specific moment in time. Since AI technologies and their adoption in UX practice are evolving rapidly, attitudes toward AI’s role in design may shift as new tools emerge and practitioners gain more experience using them. Fourth, while our approach captures a broad range of practitioner perspectives, it does not account for potential gaps in awareness regarding the historical challenges of automation. Our data collection relied on written reflections from UX professionals, but it remains unclear whether they explicitly recognize the parallels between AI-assisted design and historical automation ironies. A more targeted study, such as surveys or interviews, could help determine the extent to which practitioners are aware of these issues.

In future research, we plan to address these limitations by conducting direct interviews and surveys with UX practitioners. This would allow us to gain deeper insights into their reasoning, challenge underlying assumptions, and explore how their perspectives evolve over time as AI becomes further integrated into design workflows. Additionally, longitudinal studies tracking the long-term impact of AI on UX practice could help assess how AI adoption affects skills, creativity, and decision-making in the field.

\section{Conclusion}

As AI continues to shape UX design, it is essential to consider not just its efficiencies but also its unintended consequences. While AI automates routine tasks and aids creativity, our findings highlight concerns about de-skilling, cognitive offloading, and misplaced human responsibilities---challenges that have historically accompanied automation. UX practitioners may not fully anticipate the long-term trade-offs of AI reliance, particularly the risk of losing essential creative and cognitive skills. The persistent `substitution myth' assumes AI can replace human tasks without altering workflows, yet history shows that automation often shifts responsibilities in unforeseen ways. Rather than viewing AI as a simple tool for efficiency, it should be seen as a collaborator that requires intentional design to preserve human creativity and judgment. Future research and tool development should carefully consider these lessons from history to avoid the creation of new ironies in AI-assisted design.

\begin{acks}
This work was supported by NSF award \#2146228.
\end{acks}

\bibliographystyle{ACM-Reference-Format}
\bibliography{references}

\appendix

\end{document}